\documentclass[10pt,journal = jctcce, manuscript=reprint]{achemso}
\usepackage{lmodern}
\usepackage{lmodern}
\usepackage[latin9]{inputenc}
\usepackage{multirow}
\usepackage{amsmath}
\usepackage{amssymb}
\usepackage{graphicx}
\PassOptionsToPackage{version=3}{mhchem}
\usepackage{mhchem}
\usepackage[unicode=true,pdfusetitle,
 bookmarks=true,bookmarksnumbered=false,bookmarksopen=false,
 breaklinks=false,pdfborder={0 0 1},backref=false,colorlinks=false]
 {hyperref}

\makeatletter

\title{Relativistic semistochastic heat-bath configuration interaction}

\author{Xubo Wang}

\affiliation{Department of Chemistry, University of Colorado Boulder, Boulder,
CO 80309}

\email{xubo.wang@colorado.edu}

\author{Sandeep Sharma}

\affiliation{Department of Chemistry, University of Colorado Boulder, Boulder,
CO 80309}

\email{sandeep.sharma@colorado.edu}

\newcommand{\lyxmathsym}[1]{\ifmmode\begingroup\def\b@ld{bold}
  \text{\ifx\math@version\b@ld\bfseries\fi#1}\endgroup\else#1\fi}

\providecommand{\tabularnewline}{\\}
\usepackage{tikz}
\usetikzlibrary{calc}

\usepackage[version=3]{mhchem}

\usepackage{siunitx}
\def\H{\hat{H}}

\def\V{\mathcal{V}}

\renewcommand{\b}[1]{\ensuremath{\mathbf{#1}}}
\renewcommand{\vec}[1]{\ensuremath{\mathbf{#1}}}
\newcommand{\B}[1]{\ensuremath{\pmb #1}}

\newcommand{\bracket}[1]{\left\langle #1 \right\rangle}
\newcommand{\ket}[1]{\left|  #1  \right\rangle}
\newcommand{\wavenumber}{$\mathrm{cm^{-1}}$}

\makeatother

\begin{document}
\begin{abstract}
In this work we present the extension of semistochastic heat-bath
configuration interaction (SHCI) to work with any two-component and
four-component Hamiltonian. Vertical detachment energy (VDE) of \ce{AuH2-}
and zero-field splitting (ZFS) of \ce{NpO2^2+} are calculated by
correlating more than 100 spinors in both cases. This work demonstrates
the capability of SHCI to treat problems where both relativistic effect
and electron correlation are important.
\end{abstract}

\section{Introduction}

Relativistic effect becomes more important as one goes down the periodic
table \cite{2012_pyykko_pyykko_physicschemistryperiodic_Chem.Rev.}
\cite{2012_pyykko_pyykko_relativisticeffectschemistry_Annu.Rev.Phys.Chem.}
\cite{2012_autschbach_autschbach_perspectiverelativisticeffects_J.Chem.Phys.}
and gives rise to various phenomena, such as lanthanide contraction,
mercury being liquid, simple cubic structure of polonium etc. A proper
relativistic Hamiltonian is needed to address these effects. In practice,
the most rigorous relativistic Hamiltonian is the four-component Dirac-Coulomb-Breit
(DCB) Hamiltonian which reduces to the Dirac-Coulomb (DC) Hamiltonian
when the Gaunt term and gauge term are omitted\cite{2007_dyall_faegri_introductionrelativisticquantum_,2014_reiher_wolf_relativisticquantumchemistry_}.
The four-component Hamiltonian supports the electron-positron pair-creation
processes, however such a process involves high energy (of the order
of $2m_{e}c^{2}$ or \SI{1.02}{\mega\eV})\cite{2010_liu_liu_ideasrelativisticquantum_Mol.Phys.}
and thus does not play an important role in chemical process. In light
of this, various electron-only two-component Hamiltonians are derived
to reduce the dimension of the problem, examples of which include
the Zeroth Order Regular Approximation (ZORA) \cite{1993_lenthe_snijders_relativisticregulartwo_J.Chem.Phys.},
the Douglas-Kroll-Hess (DKH) Hamiltonian \cite{1974_douglas_kroll_quantumelectrodynamicalcorrections_AnnalsofPhysics,1986_hess_hess_Relativisticelectronicstructurecalculations_Phys.Rev.A},
the Barysz-Sadlej-Snijders (BSS) Hamiltonian \cite{2002_barysz_sadlej_infiniteordertwocomponenttheory_J.Chem.Phys.}
and exact two-component (X2C) \cite{2005_kutzelnigg_liu_quasirelativistictheoryequivalent_J.Chem.Phys.,2007_ilias_saue_infiniteordertwocomponentrelativistic_J.Chem.Phys.,2009_liu_peng_exacttwocomponenthamiltonians_J.Chem.Phys.}.
In X2C theory, the one electron operator is solved, the transformation
is then derived from the one electron solution in one step. The X2C
transformation has therefore become popular due to its efficiency
and accuracy.

Often, the one electron operator is chosen to be the one-electron
Dirac operator, and the two-electron term is simply the Coulomb operator.
By doing so, the part of spin-orbit coupling (SOC) that originates
from two-electron terms are completely neglected. One cost-effective
way to treat the two-electron SOC terms is the spin-orbit mean field
(SOMF) approach \cite{1996_hess_gropen_meanfieldspinorbitmethod_ChemicalPhysicsLetters,1996_marian_wahlgren_newmeanfieldecpbased_Chem.Phys.Lett.},
where the relativistic two-body terms are treated approximately by
including them in the Fock type one-body operator. In a SOMF calculation,
one does a non-relativistic or scalar relativistic calculation, then
computes the two electron SO integrals in the molecular orbitals obtained.
The two electron SO integrals are then contracted with the spin-averaged
self-consistent field (SCF) density matrix to obtain the effective
one-electron SO integrals. But for heavy elements, the use of scalar
orbital would cause noticeable errors\cite{2001_ilias_schimmelpfennig_inclusionmeanfieldspin_J.Chem.Phys.},
and using spinors can give more accurate results since the molecular
spinors are then fully relaxed under SOC. Such a molecular mean-field
approach uses the density matrix and spinors from a four component
mean field wave function\cite{2009_sikkema_ilias_molecularmeanfieldapproach_TheJournalofChemicalPhysics}.
In this scheme, one first does a four-component mean field calculation,
then block diagonalizes the so-obtained Fock matrix, the decoupling
of the Fock matrix is just an X2C transformation of the Fock matrix,
and is thus called X2CMMF\cite{2009_sikkema_ilias_molecularmeanfieldapproach_TheJournalofChemicalPhysics}
(exact 2-component molecular mean field). One can avoid the cost of
performing a full 4-component mean field calculation by exploiting
the local nature of the SOC, and this strategy yields the atomic mean
field (AMF) approach\cite{1996_schimmelpfennig_schimmelpfennig_amfiatomicmeanfield_,2012_li_liu_spinseparationalgebraic_J.Chem.Phys.,2018_liu_cheng_atomicmeanfieldspinorbit_TheJournalofChemicalPhysics,2020_zhang_cheng_performanceatomicmeanfield_MolecularPhysics,2022_knecht_saue_exacttwocomponenthamiltonians_}.
In AMF, an atomic SCF is performed and then the SO integral is contracted
to the mean field form. The use of atomic integrals greatly reduced
the cost and has been shown to be highly accurate. AMF approach using
4c Hamiltonian to generate mean field SO integral\cite{2018_liu_cheng_atomicmeanfieldspinorbit_TheJournalofChemicalPhysics}
including even Gaunt or Breit term has been proposed recently\cite{2022_zhang_cheng_atomicmeanfieldapproach_J.Phys.Chem.Aa,2022_knecht_saue_exacttwocomponenthamiltonians_}
and is used in most calculations in this work.

During the past few decades, quantum chemistry algorithms that have
been successful for nonrelativistic systems have also become available
for 4c and 2c Hamiltonians, such as self-consistent field\cite{1999_saue_jensen_quaternionsymmetryrelativistic_TheJournalofChemicalPhysics,2019_anderson_beylkin_diracfockcalculationsmolecules_J.Chem.Phys.},
density functional theory\cite{1997_liu_dolg_beijingfourcomponentdensity_TheorChemActa}\cite{2002_saue_helgaker_fourcomponentrelativistickohnsham_J.Comput.Chem.},
coupled cluster\cite{2001_visscher_kaldor_formulationimplementationrelativistic_J.Chem.Phys.,2010_nataraj_visscher_generalimplementationrelativistic_J.Chem.Phys.,2019_asthana_cheng_exacttwocomponentequationofmotion_J.Chem.Phys.,2019_cheng_cheng_studynoniterativetriples_J.Chem.Phys.,2021_liu_cheng_relativisticcoupledclusterequationofmotion_WIREsComput.Mol.Sci.},
configuration interaction\cite{2001_fleig_marian_generalizedactivespace_TheJournalofChemicalPhysics,2010_knecht_fleig_largescaleparallelconfiguration_J.Chem.Phys.},
multiconfiguration self-consistent field\cite{1996_jo/rgenaa.jensen_faegri_relativisticfourcomponent_J.Chem.Phys.,2008_thyssen_jensen_directrelativisticfourcomponent_J.Chem.Phys.,2015_bates_shiozaki_fullyrelativisticcomplete_J.Chem.Phys.,2018_reynolds_shiozaki_largescalerelativisticcomplete_TheJournalofChemicalPhysics,2019_jenkins_li_variationalrelativistictwocomponent_J.Chem.TheoryComput.},
multireference perturbation theory (MRPT)\cite{2015_shiozaki_mizukami_relativisticinternallycontracted_J.Chem.TheoryComput.,2022_lu_li_exacttwocomponentrelativisticmultireference_J.Chem.TheoryComput.},
multireference configuration interaction (MRCI)\cite{2015_shiozaki_mizukami_relativisticinternallycontracted_J.Chem.TheoryComput.,2020_hu_li_relativistictwocomponentmultireference_J.Chem.TheoryComput.}
etc. Several quantum chemistry packages are also available to perform
these relativistic electronic structure calculations\cite{2013_jonsson_grant_newversiongrasp2k_ComputerPhysicsCommunications,2020_saue_vanstralen_diraccoderelativistic_J.Chem.Phys.,2020_sun_chan_recentdevelopmentspyscf_J.Chem.Phys.,2018_shiozaki_shiozaki_bagelbrilliantlyadvanced_WileyInterdiscip.Rev.Comput.Mol.Sci.,2020_williams-young_li_chronusquantumsoftware_WIREsComputMolSci,2020_matthews_stanton_coupledclustertechniquescomputational_J.Chem.Phys.,2020_belpassi_storchi_berthaimplementationfourcomponent_ETJ.,2020_zhang_liu_bdfrelativisticelectronic_J.Chem.Phys.}.

In this work, we extend the semistochastic heat-bath configuration
interaction (SHCI) algorithm\cite{2017_sharma_umrigar_semistochasticheatbathconfiguration_J.Chem.TheoryComput.}
to treat two- or four-component Hamiltonians with large active spaces.
In our previous work \cite{2018_mussard_sharma_onesteptreatmentspin_J.Chem.TheoryComput.}
to treat SOC using SHCI, we used scalar orbitals with the SOMF integrals.
This time we work with spinors which is expected to give a better
description of SOC at the orbital optimization level. To our knowledge,
only density matrix renormalization group (DMRG)\cite{2014_knecht_reiher_communicationfourcomponentdensity_J.Chem.Phys.,2018_battaglia_knecht_efficientrelativisticdensitymatrix_J.Chem.TheoryComput.}
and full configuration interaction quantum Monte Carlo (FCIQMC) \cite{2020_anderson_booth_fourcomponentfullconfiguration_J.Chem.Phys.}
have been implemented for 2c/4c Hamiltonians with the capability of
treating around 100 molecular spinors, however, SHCI is often faster
than both FCIQMC and DMRG for treating non-relativistic Hamiltonians
of molecules and we expect this to be the case for relativistic Hamiltonians
as well. \cite{2015_shiozaki_mizukami_relativisticinternallycontracted_J.Chem.TheoryComput.}
This paper is organized as follows. In Section \ref{sec:Recap-of-SHCI}
and Section \ref{sec:SHCI-in-spinor}, we describe the current SHCI
algorithm and the adaptation of SHCI algorithm to treat 2c/4c Hamiltonians
including some implementation details. In Section \ref{sec:Relativistic-Hamiltonian},
we present the relativistic Hamiltonians we use in the calculations.
In Section \ref{sec:Results}, we give computational details and results
on the vertical detachment energy (VDE) of \ce{AuH2-} and the first
few excited states of \ce{NpO2^2+}.

\section{Recap of SHCI\label{sec:Recap-of-SHCI}}

\label{section 2} Semistochastic heat-bath configuration interaction
is a recently developed variant of the class of methods that perform
a selected configuration interaction followed by perturbation theory
(SCI-PT). Similar to all other SCI-PT methods\cite{1983_evangelisti_malrieu_convergenceimprovedcipsi_Chem.Phys.,2016_liu_hoffmann_iciiterativeci_J.Chem.TheoryComput.,2016_schriber_evangelista_communicationadaptiveconfiguration_J.Chem.Phys.,2020_tubman_whaley_modernapproachesexact_J.Chem.TheoryComput.},
it consists of a variational step and a perturbative step. In the
variational step, a set of important determinants is iteratively selected
by the heat-bath algorithm and the subspace eigenvalue problem is
solved. In the perturbative step, the previously obtained variational
energy is corrected by Epstein-Nesbet perturbation theory\cite{1926_epstein_epstein_starkeffectpoint_Phys.Rev.,1955_nesbet_nesbet_configurationinteractionorbital_ProcRSocLond.SerA}
to estimate the FCI energy. A semistochastic\cite{2017_sharma_umrigar_semistochasticheatbathconfiguration_J.Chem.TheoryComput.}
scheme is utilized to reduce the cost. In this section, index $i$
and $a$ represent determinants inside or outside of the current variational
space.

\subsection{Heat-bath sampling}

Given a set of initial determinants, the multi reference wave function
\begin{equation}
|\Psi\rangle=\sum_{D_{i}\in\V}c_{i}|D_{i}\rangle
\end{equation}
is obtained by diagonalizing the Hamiltonian in the current space
$\V$ of important determinants. Then new determinants that satisfy
the heat-bath criterion 
\begin{equation}
\max_{D_{i}\in\mathcal{V}}|H_{ai}c_{i}|>\epsilon_{1}\label{heat-bath}
\end{equation}
are added to the space $V$. Here $H_{ai}=\langle D_{a}|\H|D_{i}\rangle$
is the Hamiltonian matrix element and $\epsilon_{1}$ is a user defined
parameter and is usually set to as small as possible. The HCI criterion
is different from the one used in CIPSI\cite{1983_evangelisti_malrieu_convergenceimprovedcipsi_Chem.Phys.},
which is based on the contribution of a determinant $D_{a}$to the
perturbative correction to the wave function

\begin{equation}
\dfrac{\sum_{|D_{i}\rangle\in\V}H_{ai}c_{i}}{E_{0}-E_{a}}>\epsilon_{1}
\end{equation}

Although the HCI criterion is not optimal at picking out the important
determinants, the variational space formed by the two methods are
still nearly the same\cite{2017_holmes_sharma_excitedstatesusing_TheJournalofChemicalPhysics}.
Moreover, this \emph{inexpensive to evaluate }selection criterion
is implemented even more efficiently by avoiding generation of the
determinants that do not meet the criterion, speeding up both variational
and perturbative stage of the algorithm. For more detailed discussion
on efficient implementation of SHCI, we refer the readers to previous
works\cite{2017_sharma_umrigar_semistochasticheatbathconfiguration_J.Chem.TheoryComput.,2018_li_umrigar_fastsemistochasticheatbath_J.Chem.Phys.}.

\subsection{Stochastic Perturbation Theory}

After the variational stage, a perturbative step is performed to estimate
the FCI energy by Epstein-Nesbet perturbation theory, 
\begin{equation}
E_{2}=\sum_{|D_{a}\rangle\in\mathcal{C}}\frac{1}{E_{0}-E_{a}}\left(\sum_{|D_{i}\rangle\in\mathcal{V}}H_{ai}c_{i}\right)\label{pt}
\end{equation}
where $\mathcal{C}$ denotes the set of determinants that are connected
to at least one determinant in $\mathcal{V}$ by a non-zero Hamiltonian
matrix element. Since the vast majority of terms in the double sum
contribute negligibly, they can be discarded without significant loss
of accuracy and the perturbative correction can be approximated by
a ``screened sum'', 
\begin{equation}
E_{2}(\epsilon_{2})=\sum_{|D_{a}\rangle}\frac{1}{E_{0}-E_{a}}\left(\sum^{(\epsilon_{2})}H_{ai}c_{i}\right)
\end{equation}
where $\sum^{\epsilon_{2}}H_{ai}c_{i}$ includes only terms with $|H_{ai}c_{i}|>\epsilon_{2}$
and the outer sum is over the determinants $|D_{a}\rangle$ that meet
this criterion. In order to achieve good accuracy, this parameter
$\epsilon_{2}$ has to be small. Thus even with a ``screened sum'',
we might still encounter a memory bottleneck of having to store all
the determinants $\ket{D_{a}}$ and their perturbative contribution
in memory.

To further reduce the memory cost, we utilize the semistochastic perturbation
theory to estimate the perturbative correction. In our semistochastic
perturbation approach, a deterministic perturbative calculation with
a relatively loose $\epsilon_{2}^{d}$ is performed first (termed
as $E_{2}^{D}(\epsilon_{2}^{d})$). The error caused by this loose
parameter is then corrected stochastically by a much tighter $\epsilon_{2}$.
In this step, a few tens to hundreds of variational determinants are
sampled, and the perturbative correction is calculated using both
$\epsilon_{2}^{d}$ and $\epsilon_{2}$ (termed as $E_{2}^{S}(\epsilon_{2})$
and $E_{2}^{S}(\epsilon_{2}^{d})$), the final perturbation correction
is estimated by 
\begin{equation}
E_{2}(\epsilon_{2})=E_{2}^{D}(\epsilon_{2}^{d})+[E_{2}^{S}(\epsilon_{2})-E_{2}^{S}(\epsilon_{2}^{d})]
\end{equation}
The key point to this scheme is that $E_{2}^{S}(\epsilon_{2})$ and
$E_{2}^{S}(\epsilon_{2}^{d})$ are calculated using the exact same
set of determinants and thus reduce the stochastic error significantly,
with almost no increase in memory or computer time.

\subsection{Excited states}

With the four-component Hamiltonian being used, spin-orbit coupling
and all other relativistic effects are taken into account naturally,
thus $\bracket{S_{z}}$ is no longer a good quantum number, the wave
function becomes eigenfunction of the total angular momentum $J$.
The $2S+1$ fold degeneracy of a spin multiplet thus breaks into several
sets of states corresponding to different $J$ values with $2J+1$
fold degeneracy. This energy splitting between different $J$ states
can be measured experimentally to determine the zero-field splitting
(ZFS), and they are usually very small compared to the absolute energies
of the molecule.

In order to compute these excited states, the heat-bath criterion
needs to be modified. It has been done in two different ways previously\cite{2017_holmes_sharma_excitedstatesusing_TheJournalofChemicalPhysics,2018_mussard_sharma_onesteptreatmentspin_J.Chem.TheoryComput.},
the first is to replace the ground state CI vector by the maximum
values among all CI vectors 
\begin{equation}
\max_{D_{i}\in\mathcal{V}}|H_{ai}|\max_{s\in\mathrm{states}}|c_{i}^{(s)}|>\epsilon_{1}
\end{equation}
the second way is to use the averaged CI vector 
\begin{equation}
\max_{D_{i}\in\mathcal{V}}\left|H_{ai}\dfrac{\sqrt{\sum\limits _{s\in\mathrm{states}}|c_{i}^{s}|^{2}}}{\mathrm{\#\ of\ states}}\right|>\epsilon_{1}\label{sa-hci}
\end{equation}
As has been discussed before\cite{2018_mussard_sharma_onesteptreatmentspin_J.Chem.TheoryComput.},
the second way is more appropriate for handling the small splitting
and near degeneracy originating from the relativistic Hamiltonian.
We thus use equation \ref{sa-hci} as the heat-bath criterion for
excited states calculations.

\section{SHCI in spinor basis\label{sec:SHCI-in-spinor}}

\label{section 3} In this work, our relativistic SHCI implementation
works on complex-valued spinor reference wave function, any 2c or
4c Hamiltonian that works in a spinor basis can readily be used. With
the 2c/4c Hamiltonian being used, the spin symmetry and point group
symmetry which are commonly used in nonrelativistic calculations no
longer holds, we instead have Kramer symmetry and double group symmetry
that works for four-component Hamiltonian. In the current implementation,
we don't assume symmetry between barred and unbarred spinors, only
permutation symmetry and complex-conjugated symmetry is utilized.
Here, $p,q,r,s$ denote general molecular spinor index.
\begin{equation}
(pq|rs)=(rs|pq)
\end{equation}
\begin{equation}
(pq|rs)=(qp|sr)^{*}
\end{equation}

With the electron integrals being complex, the resulting Hamiltonian
matrix at the variational stage is also a complex-valued Hermitian
matrix, a complex version of the Davidson algorithm has been implemented
in our previous work to solve the complex-valued eigenvalue problem.
However, the heat-bath criterion in Equation \ref{heat-bath} and
its variant for multiple roots in Equation \ref{sa-hci} as well as
the perturbation correction in Equation \ref{pt} are not influenced
by the complex nature of the Hamiltonian since they both use the magnitude
of $H_{ai}c_{i}$. %To treat multiple roots in a single calculation, an updated HCI criterion proposed in our previous work 
%\cite{2018_mussard_sharma_onesteptreatmentspin_J.Chem.TheoryComput.} is used. It uses an averaged CI coefficient putting all roots on an equal footing.

%\begin{equation} \label{sa-hci}
%    \max_{D_i\in\mathcal{V}}\left|H_{ai}\dfrac{\sqrt{\sum\limits_{s\in \mathrm{states}}|c_i^s|^2}}{\mathrm{\#\ of\ states}}\right|>\epsilon_1 \\
%\end{equation}

\subsection{Implementation}

Here we briefly describe the implementation and the steps of leading
order cost of the algorithm. There are three major operations during
the variational stage: identify the important determinants, construct
the Hamiltonian matrix and diagonalize the matrix. In the current
implementation, all the nonzero elements of the Hamiltonian are stored
in memory using a \textit{list of lists} (LIL) format. In the LIL
format, we store a list of column index and a list of corresponding
nonzero Hamiltonian matrix elements for a given determinant. The determinants
are stored in a list of bit-packed strings that represent the occupation
of the active molecular spinors. Since spin in no longer a good quantum
number, the use of auxiliary lists implemented in nonrelativistic
case becomes complicated and more auxiliary lists are required. To
simplify the problem, we noticed that if two determinants are connected
by a single or double excitation, then there exists a determinant
with $N-2$ electrons occupied associated with both determinants.
This idea can date back to Harrison et al's full configuration interaction
(FCI) implementation\cite{1989_harrison_zarrabian_efficientimplementationfullci_ChemicalPhysicsLetters}
and used in the relativistic FCI implementation by Bates et al\cite{2015_bates_shiozaki_fullyrelativisticcomplete_J.Chem.Phys.}.
We can further say that, if a set of determinants are associated with
the same $N-2$ determinant, then they all have non-zero Hamiltonian
matrix element with each other according to the Slater-Condon rule.
To make use of this fact, we generate a list of all the $N-2$ determinants,
and then record all the $N$ determinants associated with each $N-2$
determinants. When constructing the Hamiltonian, we can make use of
these two lists to help us find the connected determinants rather
than searching the entire variation space. Also at each step, only
the matrix elements associated with the newly added determinants are
handled instead of constructing the Hamiltonian matrix from scratch
every step.

Once the Hamiltonian is generated, a complex version of Davidson algorithm
is used to obtain the lowest few eigenvalues where the most expensive
step is Hamiltonian wave function multiplication which scales as $O(kN_{V})$,
where $k$ is proportional to the fourth power of the number of electrons
and is equal to the number of columns of the Hamiltonian matrix with
non-zero values for a given determinant. The Hamiltonian matrix is
currently stored in memory, and it is the largest bottleneck in a
variational calculation.

The semistochastic perturbation scheme is identical to its nonrelativistic
variant, for detailed discussion, we recommend readers follow previous
work by one of the authors\cite{,2017_sharma_umrigar_semistochasticheatbathconfiguration_J.Chem.TheoryComput.}.

\section{Relativistic Hamiltonian\label{sec:Relativistic-Hamiltonian}}

\subsection{4c Hamiltonian}

The four-component DC(B) Hamiltonian can be written as

\begin{equation}
\hat{H}=\sum_{i}\hat{h}_{D}(i)+\frac{1}{2}\sum_{i\neq j}\hat{g}(i,j)+V_{NN}
\end{equation}
\begin{equation}
\hat{h}_{D}(i)=c^{2}(\beta-\B I_{4})+c(\B{\alpha}_{i}\cdot\b{\hat{p}}_{i})-\sum_{A}^{\text{atoms}}\dfrac{Z_{A}}{r_{iA}}
\end{equation}
\begin{equation}
\hat{g}(i,j)=\underbrace{\frac{1}{r_{ij}}}_{\mathrm{Coulomb}}-\underbrace{\underbrace{\frac{\B\alpha_{i}\cdot\B\alpha_{j}}{r_{ij}}}_{\mathrm{Gaunt}}+\underbrace{\left(\frac{\B\alpha_{i}\cdot\B\alpha_{j}}{2r_{ij}}-\frac{(\B\alpha_{i}\cdot\B r_{ij})(\B\alpha_{j}\cdot\B r_{ij})}{2r_{ij}^{3}}\right)}_{\mathrm{gauge}}}_{\mathrm{Breit}}
\end{equation}
where $\B\alpha=(\alpha_{x},\alpha_{y},\alpha_{z})$, $\alpha_{x},\alpha_{y},\alpha_{z}$
and $\beta$ are the $4\times4$ Dirac matrices 
\begin{equation}
\begin{matrix}\alpha_{x}=\begin{pmatrix}\B0_{2} & \sigma_{x}\\
\sigma_{x} & \B0_{2}
\end{pmatrix}, & \alpha_{y} & =\begin{pmatrix}\B0_{2} & \sigma_{y}\\
\sigma_{y} & \B0_{2}
\end{pmatrix},\\
\\
\alpha_{z}=\begin{pmatrix}\B0_{2} & \sigma_{z}\\
\sigma_{z} & \B0_{2}
\end{pmatrix}, & \beta & =\begin{pmatrix}\B I_{2} & 0_{2}\\
0_{2} & \B-I_{2}
\end{pmatrix}.
\end{matrix}
\end{equation}
$\sigma_{x}$, $\sigma_{y}$ and $\sigma_{z}$ are the Pauli matrices.

After the adoption of the no-pair approximation, the Hamiltonian in
the second quantization form becomes
\begin{equation}
\hat{H}=\sum_{pq}h_{pq}\hat{a}_{p}^{\dagger}\hat{a}_{q}+\frac{1}{2}\sum_{pqrs}g_{pq,rs}\hat{a}_{p}^{\dagger}\hat{a}_{r}^{\dagger}\hat{a}_{s}\hat{a}_{q}
\end{equation}
where the $p,q,r$ and $s$ indices represent positive energy spinors.
This no-pair Hamiltonian is what relativistic SHCI actually uses when
doing a four-component correlation calculation. 

\subsection{X2CAMF Hamiltonian}

In this work we use an X2CAMF Hamiltonian proposed by Liu \textit{et
al} \cite{2018_liu_cheng_atomicmeanfieldspinorbit_TheJournalofChemicalPhysics}to
treat spin-orbit coupling. In a standard X2C calculation, the bare
Coulomb term is used as the two electron operator. The spin-dependent
Coulomb interaction and the entire Breit term are then missing. To
overcome this shortcoming the molecular mean field approach\cite{2009_sikkema_ilias_molecularmeanfieldapproach_TheJournalofChemicalPhysics}
was previously proposed, whereby, a 4c Dirac Hartree Fock calculation
is performed on the entire molecule and then X2C transformation of
the entire Fock matrix is carried out. All the non-Coulomb terms as
well as the spin-orbit part of the Coulomb terms are absorbed into
the one-body operator in a mean-field fashion. The disadvantage is
that one still needs to do a molecular 4c calculation which is usually
expensive. In the AMF approach, only atomic calculations are performed
and then a separate X2C transformation for each atom is carried out.
This approach relies on the local nature of the spin-orbit coupling.
By doing atomic calculation, one can also exploit the high symmetry
in atoms which can greatly reduces the cost. A recent work by Zhang
et al\cite{2022_zhang_cheng_atomicmeanfieldapproach_J.Phys.Chem.Aa}utlizes
this feature and is implemented for the full DCB Hamiltonian.

\paragraph{AMF Hamiltonian}

We start from the spin-separation scheme for the DC Hamiltonian to
derive the atomic mean-field (AMF) Hamiltonian, the Coulomb operator
can be partitioned into a spin-free part and a spin-dependent part.
\begin{equation}
g_{pq,rs}^{\mathrm{C}}=g_{pq,rs}^{\mathrm{C,SF}}+g_{pq,rs}^{\mathrm{C,SD}}
\end{equation}
The full DCB Hamiltonian can thus be regrouped as 
\begin{equation}
\begin{split}H^{\text{DCB}}= & \sum_{pq}h_{pq}a_{p}^{\dagger}a_{q}+\frac{1}{2}\sum_{pqrs}\left(g_{pq,rs}^{\text{C,SD}}+g_{pq,rs}^{\text{Breit}}\right)\hat{a}_{p}^{\dagger}\hat{a}_{r}^{\dagger}\hat{a}_{s}\hat{a}_{q}\\
 & +\frac{1}{2}\sum_{pqrs}g_{pq,rs}^{\text{C,SF}}\hat{a}_{p}^{\dagger}\hat{a}_{r}^{\dagger}\hat{a}_{s}\hat{a}_{q}
\end{split}
\end{equation}
Although the Breit term can alse be split into spin-free term and
spin-dependent term \cite{1994_dyall_dyall_exactseparationspin_J.Chem.Phys.},
but it is still grouped together with the spin-dependent Coulomb term
and treated within the atomic mean-field approximation. The atomic
mean-field approximation is then introduced to treat the spin dependent
Coulomb and Breit term 
\begin{equation}
g_{pq}^{\mathrm{AMF}}=\sum_{A}\sum_{i_{A}\in A}n_{iA}(g_{pi_{A},qi_{A}}^{\mathrm{C,SD,A}}+g_{pi_{A},qi_{A}}^{\textrm{Breit,A}}-g_{pq,i_{A}i_{A}}^{\mathrm{C,SD,A}}-g_{pq,i_{A}i_{A}}^{\mathrm{Breit,A}})
\end{equation}
where the superscript A denotes the integral on each atom, $i_{A}$
and $n_{i_{A}}$ denote an occupied molecular orbital and the corresponding
occupation number. The DCB Hamiltonian can then be approximated as
\begin{equation}
H^{\mathrm{DCB,AMF}}\approx\sum_{pq}(h_{pq}+g_{pq}^{\mathrm{AMF}})\hat{a}_{p}^{\dagger}\hat{a}_{q}+\dfrac{1}{2}\sum_{ijkl}g_{pq,rs}^{\mathrm{C,SF}}\hat{a}_{p}^{\dagger}\hat{a}_{r}^{\dagger}\hat{a}_{s}\hat{a}_{q}
\end{equation}

\paragraph{X2C transformation}

Now we apply the standard X2C transformation\cite{2009_liu_peng_exacttwocomponenthamiltonians_J.Chem.Phys.}
to this Hamiltonian. To derive the X2C transformation, one first replaces
the small component ($\Psi^{S}$) with the pseudo large component
($\Phi^{L}$) 
\begin{equation}
\Psi^{S}=\dfrac{\B{\sigma\cdot p}}{2c}\Phi^{L}
\end{equation}
the matrix Dirac equation can then be written in a modified form 
\begin{equation}
\begin{pmatrix}\vec{V} & \vec{T}\\
\vec{T} & \dfrac{\alpha^{2}}{4}\vec{W}-\vec{T}
\end{pmatrix}\begin{pmatrix}\Psi^{L}\\
\Phi^{L}
\end{pmatrix}=\begin{pmatrix}\vec{S} & 0\\
0 & \dfrac{\alpha^{2}}{4}\vec{T}
\end{pmatrix}\begin{pmatrix}\Psi^{L}\\
\Phi^{L}
\end{pmatrix}\B{\epsilon}\label{dirac}
\end{equation}
in which $\vec{V}$ is the potential matrix, $\vec{T}$ is the kinetic
energy matrix and $\vec{W}$ is the potential matrix for small components
$(\vec{\sigma}\cdot\vec{p})V(\vec{\sigma}\cdot\vec{p})$.

In order to decouple the large and small component, a transformation
matrix $\vec{U}$ that can block diagonalize the matrix equation is
introduced 
\begin{equation}
\vec{U}=\vec{U}_{N}\vec{U}_{D},\vec{U}_{N}=\begin{pmatrix}\vec{R}_{+}^{\dagger} & 0\\
0 & \vec{R}_{-}^{\dagger}
\end{pmatrix},\vec{U}_{D}=\begin{pmatrix}\vec{I} & \vec{X}^{\dagger}\\
\vec{\Tilde{X}^{\dagger}} & \vec{I}
\end{pmatrix}.\label{unitary}
\end{equation}
$\vec{U}_{D}$ achieves the decoupling and $\vec{U}_{N}$ renormalizes
the Hamiltonian to the nonrelativistic metric. The transformed matrix
features a block diagonal form, electronic and positronic degrees
of freedom are completely decoupled 
\begin{equation}
\vec{U}\vec{h}_{D}\vec{U}^{\dagger}=\begin{pmatrix}\vec{h}_{D}^{+} & 0\\
0 & \vec{h}_{D}^{-}
\end{pmatrix}
\end{equation}
To construct $\vec{h}_{D}^{+}$, only $\vec{R}^{+}$ and $\vec{X}$
is required, thus we introduce the $\vec{X}$ matrix that relates
$\Psi^{L}$ and $\Phi^{L}$ 
\begin{equation}
\Phi^{L}=\vec{X}\Phi^{L}
\end{equation}
and the $R^{+}$ matrix that relates decoupled electronic wave function
and the original large component wave function. 
\begin{equation}
\Psi^{L}=\vec{R}^{+}\Psi^{+}
\end{equation}
\begin{equation}
\vec{R}^{+}=(\vec{S}^{-1}\vec{\Tilde{S}})^{-1/2},\vec{\Tilde{S}}=\vec{S}+\dfrac{\alpha^{2}}{2}\vec{X}^{\dagger}\vec{T}\vec{X}.
\end{equation}
The $\vec{h}_{\mathrm{D}}^{+}$ or the $\vec{h}_{\mathrm{X2C}}$ can
be written as 
\begin{equation}
\vec{h}_{\mathrm{X2C}}=\vec{R}^{+\dagger}\{\vec{h}_{D}^{LL}+\vec{X}^{\dagger}\vec{h}_{D}^{LS}+\vec{h}_{D}^{SL}\vec{X}+\vec{X}^{\dagger}(\vec{h}_{D}^{SS})\vec{X}\}\vec{R}^{+}\label{X2C-1e}
\end{equation}
Note the above transformation only transforms the one-electron Dirac
Hamiltonian, so we call it one-electron X2C Hamiltonian ($\vec{h^{X2C-1e}}$).
The AMF term is transformed atomically. For each atom, a 4c calculation
is performed, and the atomic Fock matrix takes the place of $\vec{h}^{D}$
to determine the atomic $X$ and $R$ matrices. This atomic $X$ and
$R$ matrices are then used to transform $\vec{h^{AMF}}$ to $\vec{h^{2c,AMF}}$.
The X2CAMF Hamiltonian\cite{2022_zhang_cheng_atomicmeanfieldapproach_J.Phys.Chem.Aa}
is then written as 
\begin{equation}
\vec{H^{\mathrm{X2CAMF}}}=\sum_{ij}(h_{ij}^{\mathrm{X2C-1e}}+h_{ij}^{\mathrm{2c,AMF}})E_{ij}+\dfrac{1}{2}\sum_{ijkl}g_{ijkl}^{\mathrm{NR}}E_{ijkl}\label{x2camf}
\end{equation}
% might need to remark on this Hamiltonian.

\section{Results\label{sec:Results}}

\subsection{Computational details}

We present calculations to evaluate the photoelectron detachment energies
(DE) of \ce{AuH2-} which results in the formation of neutral open-shell
molecule \ce{AuH2} in different states. We also calculate the zero-field
splitting (ZFS) of \ce{NpO2^2+}. The X2CAMF Hamiltonian is computed
using X2CAMF package by Zhang \cite{2022_zhang_zhang_x2camf_}. The
X2CAMF Hartree-Fock (HF) calculations are performed through the \href{https://github.com/xubwa/socutils}{socutils}
code by one of the authors\cite{2022_wang_wang_xubwasocutils_}. The
SHCI calculations are performed using the ZSHCI module of the Dice
code by the authors\cite{2022_sharma_sharma_dice_}. The input and
output for all calculations can be accessed from a public repository\cite{2022_wang_wang_xubwarelhci_}.
All the SHCI calculations use $\epsilon_{2}$value of $10^{-10}$a.u.
and the stochastic errors are converged to $5\times10^{-6}$ a.u.
in order to recover the degenracy between Kramer doublets. We extrapolate
to the FCI limit by fitting the total energy $E_{tot}=E_{var}+E_{2}$
with respect to the PT2 correction $E_{2}$\cite{2017_holmes_sharma_excitedstatesusing_TheJournalofChemicalPhysics}.
The extrapolation error is estimated to be one-fifth of the difference
between the calculated energy with the smallest value of $\epsilon_{1}$.

\subsection{Vertical detachment energy of \protect\ce{AuH2-}}

\begin{figure}
\includegraphics[width=0.2\textwidth]{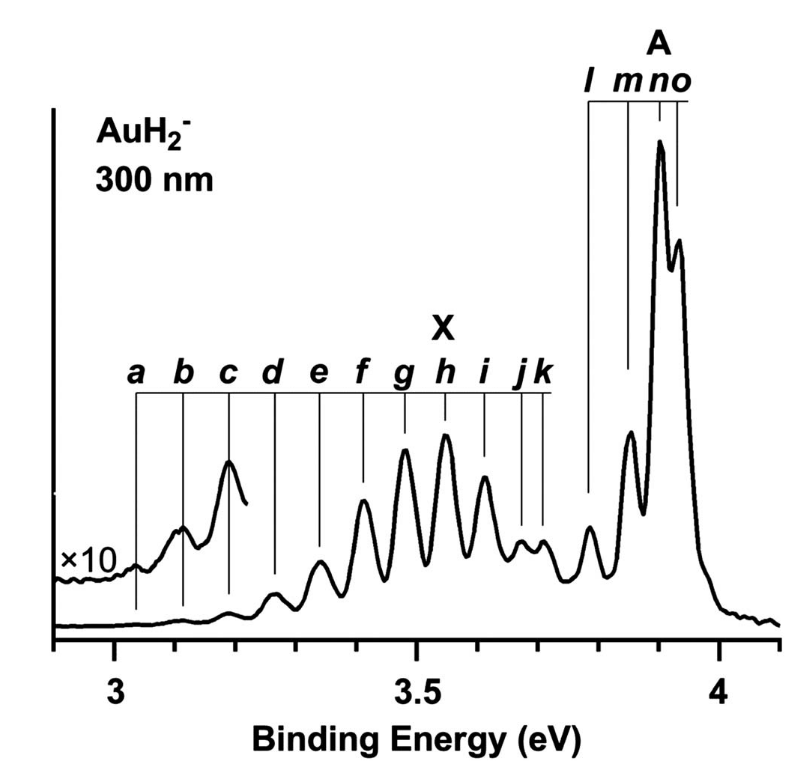}

\caption{Vibrational resolved PES spectra for X state and A state from the
experiments published in paper\cite{2012_liu_wang_electronicstructurechemical_Chem.Sci.}.\label{fig1:exp}}
\end{figure}

The photoelectron spectrum of \ce{AuH2-} was accurately measured
by Liu et al\cite{2012_liu_wang_electronicstructurechemical_Chem.Sci.}.
In the experiment work, the vibrational spectra of X state and A state
are well resolved and were discussed carefully. The X state is the
ground state of the \ce{AuH2} and it has a bent geometry, peak ``a''
to peak ``k'' are considered to be the result of vibrational progression,
the peak ``a'' corresponds to the adiabatic detachment energy (ADE),
peak ``h'' with the highest intensity is deemed to be the vertical
detachment energy (VDE). The spacing between these peaks (except for
peak ``k'') can be fit well using the anharmonic vibrational model:

\begin{equation}
E_{\nu}=(\nu+1/2)\omega_{h}-(\nu+1/2)^{2}\omega_{h}\chi_{e}
\end{equation}

Peak ``l'' to peak ``o'' are assigned to be vibration levels of
the A state, but the spacing between them do not fit into this model
well. The authors proposed a slightly bent structure for A. More recent
work by Sorbelli et al\cite{2020_sorbelli_belpassi_groundexcitedelectronic_Phys.Chem.Chem.Phys.}
used X2C-EOM-CCSD to calculate the system and reinterpret the spectrum.
They have proposed that state A has a linear structure, the unusual
behavior in the peak ``l'' to ``o'' was explained as the pseudo-Jahn-Teller
(PJTE) effect. They claimed that the PJTE induces a symmetry breaking
along the asymmetric stretching coordinate, the centrosymmetric linear
nuclear configuration thus becomes a saddle point and the most stable
configuration would be an asymmetric configuration. Here we calculated
the vertical detachment energies (VDE) from the experimentally measured
X band up to E band. The main focus in this work is not to give any
new explanation to this spectroscopy problem but rather to have a
comparison with the EOM-CCSD results especially when SHCI can give
near exact results to see how they compare. The geometry ($r_{\mathrm{Au-H}}=1.647\lyxmathsym{\AA},\alpha_{\mathrm{H-Au-H}}=180^{\circ}$)
is taken from the experiment work\cite{2012_liu_wang_electronicstructurechemical_Chem.Sci.}.
The X2CAMF-HF wave function for closed shell \ce{AuH2-} are used
for both \ce{AuH2-} and \ce{AuH2} calculation as the reference wave
function.

The VDEs from X state up to E state using EOM-CCSD with both X2CAMF
and X2CMMF Hamiltonian and SHCI with X2CAMF Hamiltonian under different
active spaces are listed in Table \ref{tab:AuH2}. Both Hamiltonians
contain relativistic effects up to the Gaunt term. The Dyall's triple
zeta basis set\cite{2009_dyall_gomes_revisedrelativisticbasis_TheorChemAcc,2016_dyall_dyall_relativisticdoublezetatriplezeta_TheorChemAcc}
is used for all atoms based on results from a previous theory paper\cite{2020_sorbelli_belpassi_groundexcitedelectronic_Phys.Chem.Chem.Phys.}.
One large EOM-CCSD calculation which correlates virtual spinors up
to 100 Hartree is also performed and is used to estimate the missing
dynamic correlation in the SHCI calculation with 124 spinors correlated
as shown in equation \ref{eq:shci_composite}. The extrapolation error
are within 0.007 eV for all 82 spinor calculations and 0016 eV for
124 spinor calculations.
\begin{equation}
\mathbf{E_{composite}^{SHCI}=E_{124\thinspace spinors}^{SHCI}-E_{124\thinspace spinors}^{EOMCC}+E_{100\thinspace Hartree}^{EOMCC}}\label{eq:shci_composite}
\end{equation}

We start by looking at the difference between X2CAMF and X2CMMF Hamiltonian
from EOM-CCSD calculations at three different active spaces. The energy
of X2CAMF is systematically lower by \SI{1}{\milli\eV} than
X2CMMF for state X and state A. For the other four states, the X2CAMF
energy is higher than X2CMMF for around \SI{4}{\milli\eV}.
This difference between the two different Hamiltonians is consistent
and also much smaller than other uncertainties in the calculations.

Then we may compare the EOM results and SHCI results with 82 spinors
and 124 spinors active space. In the 82 spinors calculation, the X
state and A state are misordered for both methods. If we compare the
VDEs for state A to state E between the methods, we notice that the
EOM-CCSD systematically underestimates them by 0.255 eV to 0.273 eV
while it only underestimates the X state by 0.088 eV. When the larger
active space with 124 spinors is used, the EOM-CCSD still underestimates
the VDEs compared to the near-exact SHCI results, but the discrepancy
reduces to within 0.130 eV to 0.156 eV. The VDE for X state however,
behaves differently. The VDE from SHCI decreases while the EOM-CCSD
VDE increases. While the composite SHCI energy gives a good agreement
with experiment for the X state and A state, the VDE from B state
to E state are all overestimated while the reference EOM results achieve
a better agreement with experiment values. This good agreement with
experiments is likely caused by some fortuitous error cancellation.
The composite VDE of X state and A state agrees with experiment value
better than the EOM-CCSD result. In particular, if one looks at the
difference between energies of the X and the A states, the EOM-CCSD
always give a smaller gap between the two states since it underestimates
the energy of the A state.

\begin{table*}[t]
\caption{VDEs (in \si{\eV}) of \protect\ce{AuH2-} with 5d and 6s electrons
correlated. The EOM results underestimate energy of state A at all
active spaces. \label{tab:AuH2}}

\begin{tabular}{cccccccccccccc}
\hline 
\multirow{1}{*}{State} & \multicolumn{3}{c}{82 spinors} &  & \multicolumn{3}{c}{124 spinors} &  & \multicolumn{3}{c}{up to 100 Hartree} &  & \multirow{1}{*}{Experiment}\tabularnewline
\hline 
%\cline{1-10} \cline{2-10} \cline{3-10} \cline{4-10} \cline{5-10} \cline{6-10} \cline{7-10} \cline{8-10} \cline{9-10} \cline{10-10} 
 & EOM  & EOM  & SHCI  &  & EOM  & EOM  & SHCI  &  & EOM  & EOM  & SHCI  &  & \tabularnewline
 & MMF  & AMF  & AMF  &  & MMF  & AMF  & AMF  &  & MMF  & AMF  & (composite)  &  & \tabularnewline
\hline 
X  & 3.517  & 3.516  & 3.604  &  & 3.545  & 3.544  & 3.521  &  & 3.666  & 3.665  & 3.641  &  & 3.678\tabularnewline
A  & 3.292  & 3.292  & 3.560  &  & 3.627  & 3.626  & 3.744  &  & 3.792  & 3.791  & 3.909  &  & 3.904\tabularnewline
B  & 4.019  & 4.023  & 4.296  &  & 4.499  & 4.504  & 4.653  &  & 4.740  & 4.745  & 4.895  &  & 4.635\tabularnewline
C  & 4.123  & 4.127  & 4.398  &  & 4.582  & 4.587  & 4.741  &  & 4.834  & 4.838  & 4.992  &  & 4.785\tabularnewline
D  & 5.128  & 5.131  & 5.372  &  & 5.510  & 5.513  & 5.643  &  & 5.768  & 5.771  & 5.902  &  & 5.745\tabularnewline
E  & 5.600  & 5.604  & 5.859  &  & 6.007  & 6.011  & 6.167  &  & 6.275  & 6.280  & 6.435  &  & 6.220\tabularnewline
\hline 
\end{tabular}
\end{table*}

\subsection{\ce{NpO2^2+}}

The rather stable actinyl ions, \ce{AnO2^n+}, as well as their derivatives
have interesting electronic structures and magnetic properties due
to the similar order of SOC and crystal field effects and have therefore
drawn people's attention. Previous work by Gendron et al \cite{2014_gendron_autschbach_magneticpropertieselectronic_Chem-EurJ,2014_gendron_autschbach_magneticresonanceproperties_Inorg.Chem.}
has systematically studied neptunyl ion \ce{NpO2^2+} and its derivative
using multireference methods and includes SOC by means of state interaction.
A state interaction version of DMRG\cite{2016_knecht_reiher_nonorthogonalstateinteractionapproach_J.Chem.TheoryComput.}
and our previous one-step SHCI treatment of SOC \cite{2018_mussard_sharma_onesteptreatmentspin_J.Chem.TheoryComput.}
have been used to calculate the energies of the neptunyl ion. It is
worth mentioning that all the previous calculations are based on different
spin-free reference wave functions, and includes the SOC terms at
correlation level.

We use uncontracted ANO-RCC basis\cite{2005_roos_widmark_newrelativisticano_J.Phys.Chem.A}
for Np and uncontracted cc-pVTZ basis\cite{1989_dunning_dunningjr_gaussianbasissets_J.Chem.Phys.}
for O. The linear geometry with both Np-O bond length are 1.70 \AA\ is
taken from the work of Gendron et al\cite{2014_gendron_autschbach_magneticpropertieselectronic_Chem-EurJ}.
A fraction occupation X2CAMF-HF is used as the reference state. The
one open-shell electron is averaged in 4 spatial orbitals to give
equal description on the four lowest Kramer doublets. The energies
of the four doublets from this work as well some previous results
are tabulated in Table \ref{tab:npo2}. The extrapolation error for
all states are within 80 \wavenumber. The spinor calculation gives
different results than the previous spin-orbit calculations, with
the splittings between $^{2}\Phi$ states and $^{2}\Delta$ states
relative to the ground state are generally smaller. The difference
can be attributed to the higher accuracy of the reference wave function
in our present X2CAMF calculations where the orbital relaxation due
to SOC is fully included while in other calculations, SOC is only
taken into account at the correlation level. Though the influence
is generally not that large for lighter elements, Np is heavy enough
so that the difference between the spinor reference and the scalar
reference can be large. 
\begin{table*}
\caption{Relative energies (\wavenumber) of the electronic states of \ce{NpO2^2+}
calculated with X2CAMF Hamiltonian with Breit term included for two
different active space. All previous calculations using scalar relativistic
orbitals underestimate the energy of $^{2}\Delta_{3/2}$ state and
overestimated the energy of $^{2}\Phi_{7/2}$ state.\label{tab:npo2}}

\centering{}%
\begin{tabular}{cccccc}
\hline 
 & \multicolumn{2}{c}{SHCI(current work)} & CASPT2-SO\cite{2014_gendron_autschbach_magneticresonanceproperties_Inorg.Chem.}  & CASSCF-SO\cite{2014_gendron_autschbach_magneticresonanceproperties_Inorg.Chem.}  & SO-SHCI\cite{2018_mussard_sharma_onesteptreatmentspin_J.Chem.TheoryComput.}\tabularnewline
State  & (4o,1e)  & (60o,13e)  & (10o,7e)  & (10o,7e)  & (143o,17e)\tabularnewline
\hline 
$^{2}\Phi_{5/2}$  & 0  & 0  & 0  & 0  & 0\tabularnewline
$^{2}\Delta_{3/2}$  & 3687  & 3429  & 3011  & 3179  & 3857\tabularnewline
$^{2}\Phi_{7/2}$  & 7640  & 7165  & 8092  & 8077  & 8675\tabularnewline
$^{2}\Delta_{5/2}$  & 9171  & 8868  & 9192  & 9288  & 10077\tabularnewline
\hline 
\end{tabular}
\end{table*}

\section{Conclusions}

We have extended the SHCI algorithm to treat general two-component
Hamiltonian for both the ground and excited states. Our calculations
show that SHCI is capable of treating relativistic Hamiltonian with
over 100 spinors. Application on VDEs of \ce{AuH2} gives a better
gap between X state and A state which outperforms EOM-CCSD at the
same basis set. The low energy spectrum of \ce{NpO2^2+} demonstrates
that for such heavy elements, it is necessary to include relativistic
effects at the SCF level.

The current method still has two limits, the variational Hamiltonian
is very memory intensive because of the need to store the Hamiltonian
and limits the variational space to a few million determinants. Thus
an efficient method to obtain the variational wave function is needed.
A matrix free eigen solver based on the coordinate descent algorithm\cite{2019_wang_lu_coordinatedescentfull_J.Chem.TheoryComput.}
is in development. Due to the number of electrons and large basis
used in a relativistic calculation, even SHCI cannot treat sufficient
number of orbitals in the active space to account for dynamical correlation.
Work in this direction is under-way, we are working on a relativistic
phaseless auxiliary field quantum Monte Carlo\cite{2018_motta_zhang_initiocomputationsmolecular_WIREsComput.Mol.Sci.,2022_eskridge_zhang_initiocalculationsatoms_J.Chem.Phys.a}
with the relativistic HCI wave function as the trial state\cite{2022_mahajan_sharma_selectedconfigurationinteraction_J.Chem.Phys.}.

\section{Acknowledgments}

XW was supported through the National Science Foundation grant CHE-2145209.
SS was supported by the grant from the Camille and Henry Dreyfus foundation.
\bibliography{revised}

\end{document}